\documentclass{pasj00}

\Received{$\langle$reception date$\rangle$}
\Accepted{$\langle$acception date$\rangle$}
\Published{$\langle$publication date$\rangle$}
\SetRunningHead{Active Region Heating}{Active Region Heating}

\usepackage{times}

\begin{document}

\title{Observations of Transient Active Region Heating with Hinode}
\author{Harry \textsc{P. Warren}\altaffilmark{1},
        Ignacio \textsc{Ugarte-Urra}\altaffilmark{1,2},
	David H. \textsc{Brooks}\altaffilmark{1,2},
	Jonathan W. \textsc{Cirtain}\altaffilmark{3},
	David R. \textsc{Williams}\altaffilmark{4}, and
	Hirohisa \textsc{Harra} \altaffilmark{5}}

\altaffiltext{1}{Space Science Division, Naval Research Laboratory,
Washington, DC 20375, USA}
\altaffiltext{2}{College of Science, George Mason University, 4400
University Drive, Fairfax, VA 22030, USA}
\altaffiltext{3}{Harvard-Smithsonian Center for Astrophysics,
60 Garden Street, Cambridge, MA 028138, USA}
\altaffiltext{4}{Mullard Space Science Laboratory, University College
London, Holmbury St Mary, Dorking, Surrey, RH5 6NT, UK}
\altaffiltext{5}{National Astronomical Observatory of Japan, Mitaka, Tokyo,
181-8588}

\email{hwarren@nrl.navy.mil, iugarte@ssd5.nrl.navy.mil}

\maketitle

\begin{abstract}

 We present observations of transient active region heating events
 observed with the Extreme Ultraviolet Imaging Spectrometer (EIS) and
 X-ray Telescope (XRT) on Hinode. This initial investigation focuses
 on NOAA active region 10940 as observed by Hinode on February 1, 2007
 between 12 and 19 UT.  In these observations we find numerous
 examples of transient heating events within the active region. The
 high spatial resolution and broad temperature coverage of these
 instruments allows us to track the evolution of coronal plasma. The
 evolution of the emission observed with XRT and EIS during these
 events is generally consistent with loops that have been heated and
 are cooling.  We have analyzed the most energetic heating event
 observed during this period, a small GOES B-class flare, in some
 detail and present some of the spectral signatures of the event, such
 as relative Doppler shifts at one of the loop footpoints and enhanced
 line widths during the rise phase of the event.  While the analysis
 of these transient events has the potential to yield insights into
 the coronal heating mechanism, these observations do not rule out the
 possibility that there is a strong steady heating level in the active
 region. Detailed statistical analysis will be required to address
 this question definitively.

\end{abstract}

\section{Introduction}

 Understanding how the solar corona is heated to high temperatures is
 one of the principal objectives of the Hinode mission
 \citep{kosugi2007}. The Extreme Ultraviolet Imaging Spectrometer
 (EIS) on Hinode has an unprecedented combination of high spatial,
 spectral, and temporal resolution and provides a wealth of
 information on the temperatures, densities, velocities, and
 nonthermal motions in the solar corona. The X-ray Telescope (XRT) on
 Hinode is one of the highest spatial resolution soft X-ray telescopes
 ever built, and combines this high spatial resolution with high
 temporal resolution and a broad temperature coverage. Together, these
 instruments give us an unparalleled ability to probe the morphology
 and evolution of coronal plasma (see, \cite{culhane2007} and
 \cite{golub2007}, for more details on the instruments).

 One question of particular importance is the time scale for heating
 in solar active regions. Recent observations have yielded
 contradictory results. At high temperatures, observations at soft
 X-ray wavelengths have suggested that the emission is relatively
 steady and can be described with steady heating models. The analysis
 of individual loops (e.g., \cite{porter1995,kano1996}) has shown that
 the evolution of the intensity is generally much slower than the
 conductive or radiative cooling times and that the loop emission is
 consistent with the RTV scaling laws \citep{rosner1978}. Transient
 brightenings have been observed in active regions previously (e.g.,
 \cite{shimizu1992}), but they do not appear to provide sufficient
 heating (e.g., \cite{shimizu1995}). The application of steady heating
 models to entire active regions (e.g., \cite{warren2006b}), coronal
 bright points \citep{brooks2007}, and the full Sun
 \citep{schrijver2004} have successfully reproduced the observed high
 temperature emission.

 \begin{figure*}[t!]
  \centerline{%
  \FigureFile(6.4in,3.2in){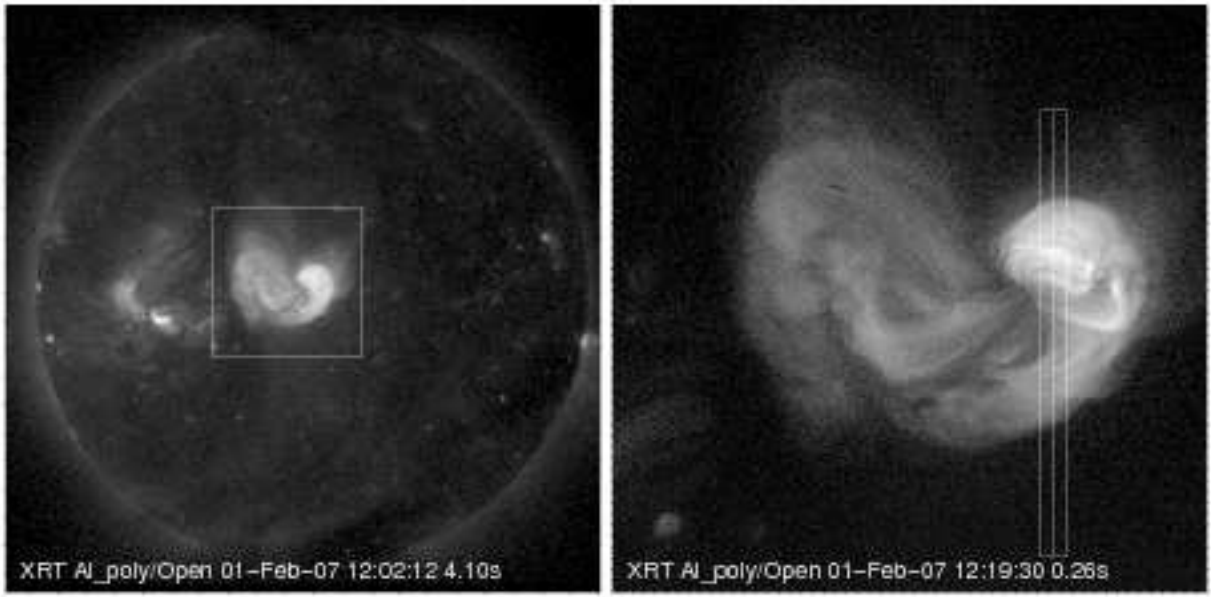}}
  \caption{XRT observations of AR 10940 on February 2, 2007. On the
  left is a composite full disk image. On the right is an example of a
  $512\times512$ pixel cutout. The field of view for the EIS context
  and sit-and-stare observations is indicated. A movie is provided as
  on-line material.}
 \label{fig:1}
 \end{figure*}

 At lower temperatures, observations of active regions at temperatures
 near 1\,MK suggest that coronal plasma is far from equilibrium. The
 analysis of coronal loops at these temperatures indicate very large
 densities relative to the RTV scaling laws (e.g.,
 \cite{winebarger2003,aschwanden2001b,lenz1999}). Full active region
 modeling shows that while the steady heating models can reproduce the
 high temperature emission, these models cannot account for the
 presence of cooler active region loops \citep{warren2006b}. Because
 loops cool more rapidly than they drain (e.g., \cite{cargill1995}),
 impulsive heating models can account for the bright active region
 loop emission observed at these temperatures (e.g.,
 \cite{warren2002b,warren2003}).  Examples of active region loops
 cooling from high temperatures down through 1\,MK have been
 documented (e.g., \cite{winebarger2005,urra2006}).

 The different heating time scales suggested by the observations at
 different temperatures can be resolved in a number of ways. It may be
 that there is a different heating mechanism responsible for the
 plasma at different temperatures. Alternatively, it may be that high
 temperature emission is more dynamic than previously thought and the
 modest spatial resolution of previous instruments has obscured
 this. The purpose of this paper is to make an initial survey of some
 EIS and XRT active region observations and to explore the available
 diagnostics. 

 \section{Observations}

 NOAA active region 10940 traversed the solar disk observable from
 Earth from January 27 to February 10, 2007. In this paper we consider
 results from specially designed EIS active region observing sequences
 taken during the period February 1--6, 2007. The EIS observations
 during this period generally consisted of small context rasters
 ($40\arcsec\times400\arcsec$) used for co-alignment with XRT that
 were followed by a long series of exposures at a fixed position with
 the 1\arcsec\ slit taken at a cadence of 30\,s. In both observing
 sequences 20 spectral windows 24 pixels wide and 400\arcsec\ long
 were selected to be included in the telemetry stream. These EIS
 sit-and-stare observations were typically run for periods of
 approximately 10 consecutive hours tracking a single position on the
 Sun.

 Representative images from the XRT are shown in
 Figure~\ref{fig:1}. The XRT observations during this time generally
 consisted of $512\times512$ pixel cutouts at a cadence of about 60\,s
 in the Al\_poly/Open filter combination. XRT images in the
 Al\_poly/Al\_thick filter combination were taken at a cadence of
 about 360\,s. Full disk synoptic observations in multiple filters
 were taken 4 times per day. Sample XRT images of this active region
 are shown in Figure~\ref{fig:1}.

\section{Data Reduction}

 \begin{figure*}[t!]
 \centerline{%
 \FigureFile(5.76in,1.917in){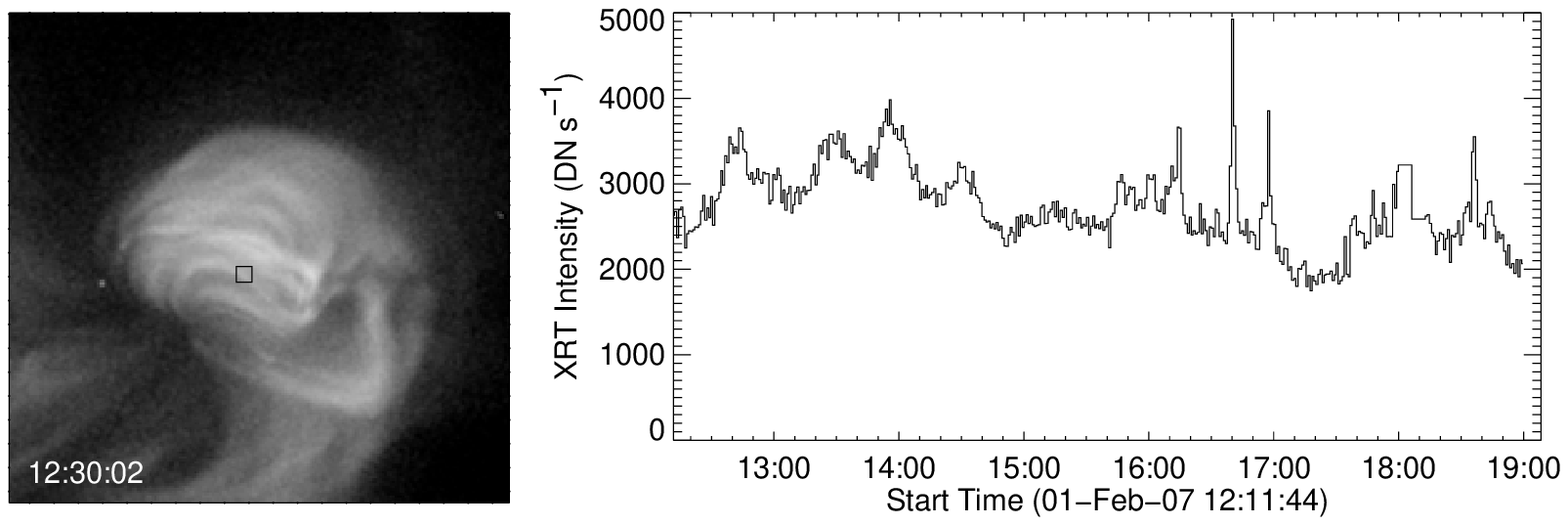}}
 \centerline{%
 \FigureFile(5.76in,1.917in){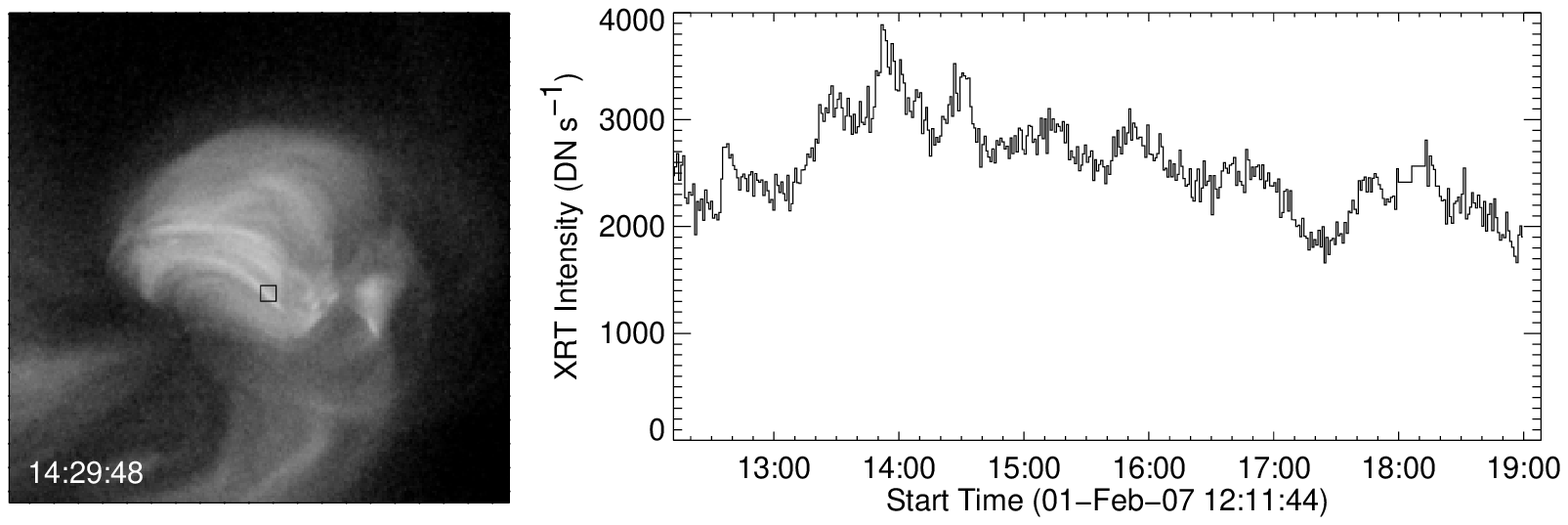}}
 \centerline{%
 \FigureFile(5.76in,1.917in){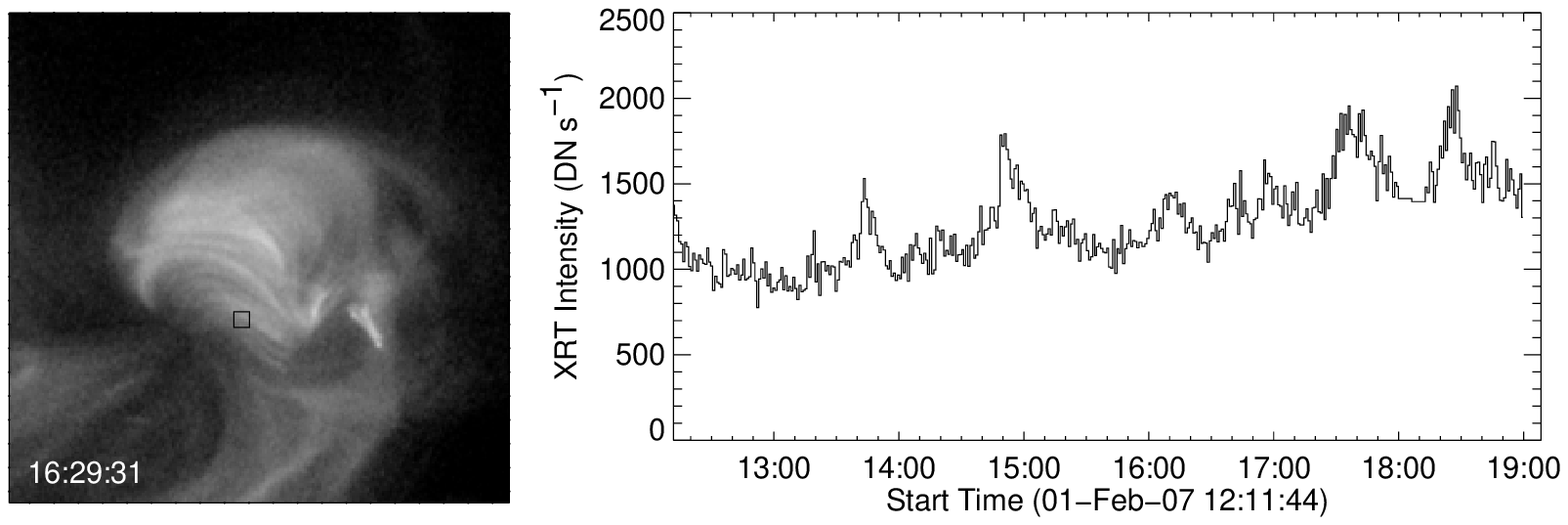}}
 \caption{Example XRT Al\_poly/Open images and light curves for the
  bright emission in AR10940. The light curves are for single pixel
  locations and are indicated by the box on the corresponding
  image. We have not estimated the uncertainties in the XRT data but
  the intensity fluctuations in successive images give some indication
  of the noise level.}
 \label{fig:2}
 \end{figure*}

 EIS spectra have been processed using current standard reduction
 software, which consists of CCD bias and dark current subtraction and
 cosmic ray and hot pixel removal. Pixels affected by the latter are
 flagged as missing and their values are replaced by the median value
 of the neighboring ones.

 Spectral lines were fitted with single Gaussian profiles, except for
 the Fe\,\textsc{xiii} doublet (203.83/203.80\,\AA) line which is a
 resolved blend with the Fe\,\textsc{xii} 203.73\,\AA\ line. The fits
 return peak intensity, width, and center position of the line for
 every pixel along the slit and every exposure. Since the spacecraft
 tracks solar rotation in the sit-and-stare observations we obtain
 temporal variations of three parameters for the same position on the
 Sun. 

 The spectral information has to be corrected of two instrumental
 effects: the tilt of the slit on the CCD and the sinusoidal drift of
 the lines on the detector due to orbital changes. Further details and
 how to do the correction is given elsewhere (Mariska et al., this
 issue).  Spatially, a North-South spacecraft drift is evident in the
 time slice intensity image of every line, i.e. the image that results
 from putting together in succession the intensity images of the slit
 for every exposure.  This jitter is also present in the XRT movies,
 and can be corrected by cross-correlation of the XRT images with
 respect to a reference image, which was chosen to be the XRT image
 closest to the middle of the EIS context raster. The amplitudes of
 these offsets, which range from 1 to 4 XRT pixels, are used in first
 order to correct the spacecraft drift along the slit. The correction
 is satisfactory for the time series we have analyzed.

 The co-alignment between XRT and EIS was established by
 cross-correlating the Al\_poly/Open image taken closest to the middle
 of the EIS context raster in the Fe\,\textsc{xvi} 262.98\,\AA\ line. For
 non-flaring active region conditions the morphology of the XRT
 Al\_poly/Open images is very similar to the EIS Fe\,\textsc{xvi}
 262.98\,\AA\ raster.
 
 From the location of the EIS slit on the co-aligned XRT datacube we
 extract a time slice, i.e. a Solar Y versus time image, that can be
 directly compared with the EIS time slices described above. The
 Al\_poly XRT filter peak response is at around 8\,MK, in the upper
 limit range of EIS temperature coverage, which makes it, apart from
 an excellent context image, a good complement in terms of temperature
 sampling.

\section{Results}

 XRT movies of this active region immediately reveal the dynamic
 nature of the high temperature emission. A wide variety of loops of
 different lengths and topological connections are observed to
 brighten up and disappear. Despite the rapid evolution of many loops
 within the active region, the general morphology of the active region
 is relatively constant, and the total intensity appears to evolve
 slowly.

 To make these impressions more quantitative we have calculated light
 curves for individual pixels in the co-aligned XRT data cubes. As
 illustrated by the light curves shown in Figure~\ref{fig:2}, there
 are numerous pixels which show intensity variations of more than
 100\% during the course of the observations. The accompanying XRT
 images show that these intensity variations are related to
 brightenings in individual loops. The average intensity in the active
 region, in contrast, is much more steady, varying by only $\pm$20\%
 during the 10 hours of observation.

 \begin{figure*}[t!]
 \centerline{\FigureFile(5.5in,7.7in){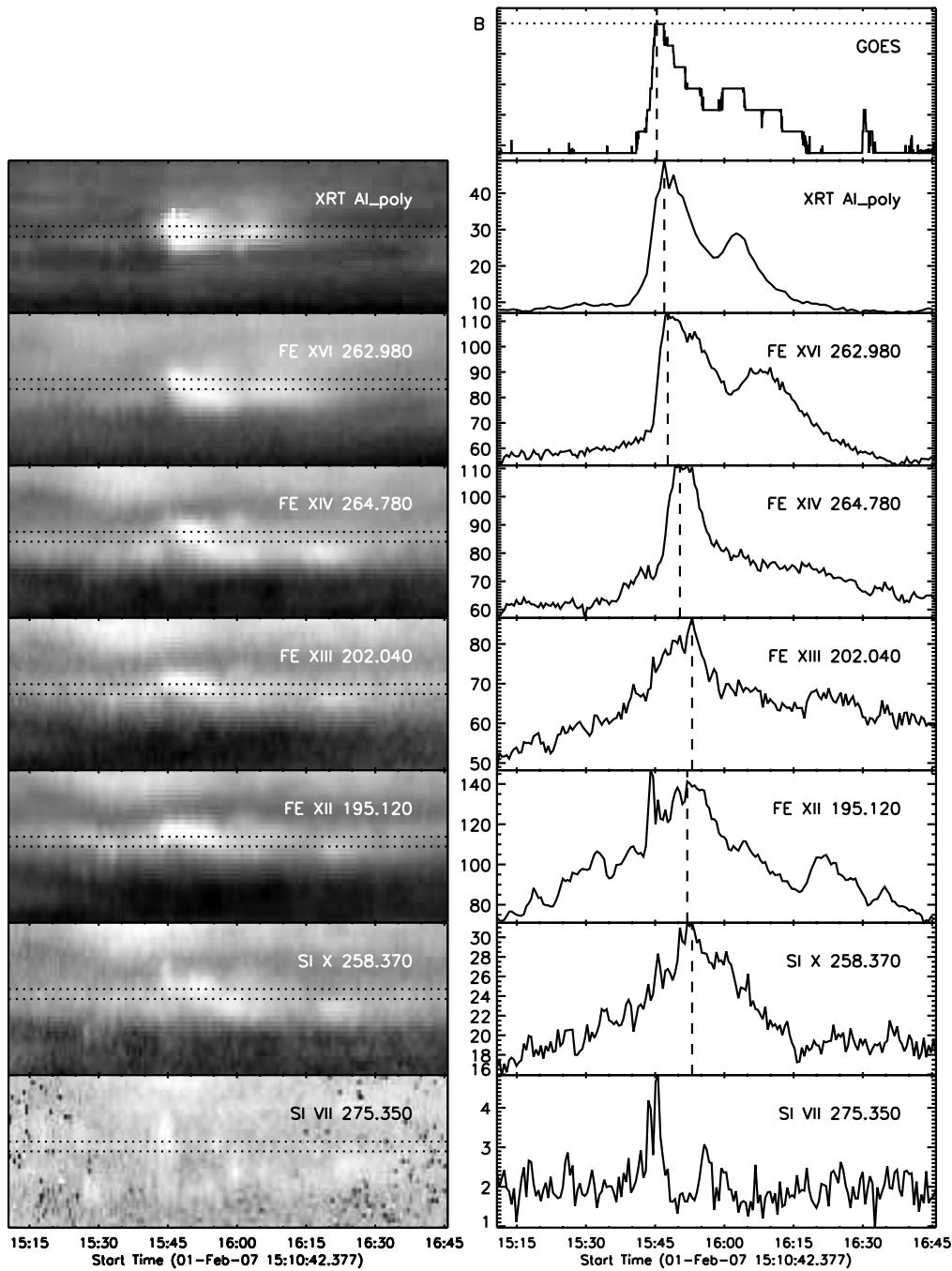}}
 \caption{Stack plots and light curves for a B flare observed on
 February 1, 2007 near 15:45 UT. On the left plots of intensity along
 the slit as a function of time are shown. On the right light curves
 are shown. The light curve has been computed from the three spatial
 pixels indicated by the dotted line. The dashed vertical line is the
 approximate position of the peak intensity. The units for the EIS
 intensities are $10^3$ ergs cm$^{-2}$ s$^{-1}$ sr$^{-1}$. The XRT
 intensities are in units of $10^3$ DN s$^{-1}$. The spatially
 integrated GOES 1--8\,\AA\ light curve is also shown.}
 \label{fig:3}
 \end{figure*}

 To aid in the identification of events observed with EIS and XRT we
 have inspected stack plots of integrated line intensity as a function
 of time. We generally begin by identifying events the XRT data and
 then look for the corresponding signatures in the EIS
 observations. We have found many examples of transient brightenings
 in these data. The general trend in the light curves supports the
 idea that these brightenings are heating events in individual
 loops. Intensity enhancements are usually first seen in the detector
 sensitive to hotter temperatures, the XRT Al\_poly filter, and then
 in successively cooler emission lines: Fe\,\textsc{xvi} (2.5 MK),
 Fe\,\textsc{xiv} (2.0 MK), Fe\,\textsc{xiii} (1.6 MK),
 Fe\,\textsc{xii} and Si\,\textsc{x} (1.3 MK), and Si\,\textsc{vii}
 (0.6 MK). While this is the general trend there are many instances
 where the evolution of the plasma is confusing, presumably because of
 multiple structures along the line of sight.

 \begin{figure*}[t!]
 \centerline{\FigureFile(6.67in,4in){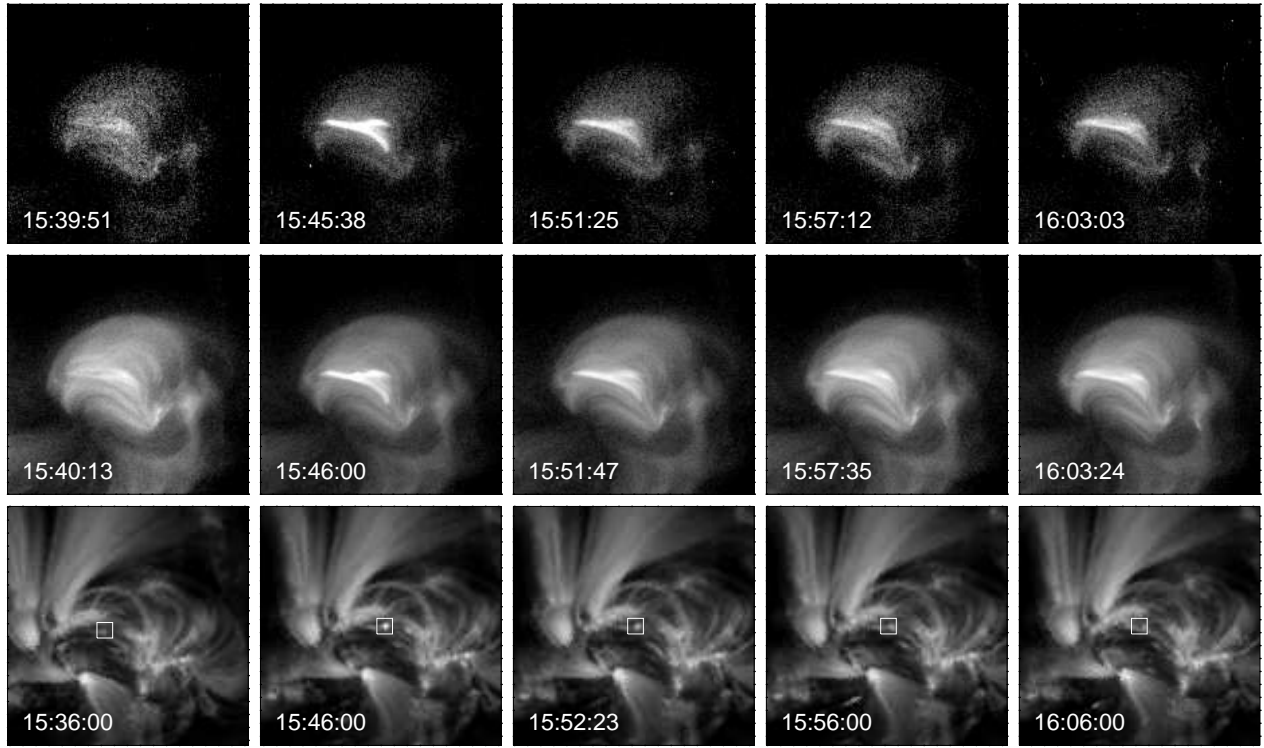}}
 \caption{XRT (Al-poly/Open and Al-poly/Al-thick) and STEREO B EIVI
 171\,\AA\ images for the B flare observed on February 1, 2007 near
 15:45. The spatial position of the footpoint spectra shown in
 Figure~\ref{fig:5} is indicated by the box. A movie is provided as
 on-line material.}
 \label{fig:4}
 \end{figure*}

 For this initial study we focus on the brightest event that has the
 clearest signatures, a GOES B-class flare observed on February 1,
 2007 near 15:45 UT. The stack plots and light curves from this event
 are shown in Figure~\ref{fig:3}. Data from only 6 of the 20 available
 EIS spectral windows are given. There is a clear progression in the
 time of the peak emission from the XRT Al\_poly to the EIS
 Si\,\textsc{x}. At the lower temperatures the signature of the event
 is not as clear, suggesting some contamination from foreground and
 background emission. The Si\,\textsc{vii} light curve is particularly
 difficult to understand. The emission in this line peaks very early,
 before an significant rise in the high temperature light curves and
 there is very little emission in Si\,\textsc{vii} during the decay of
 the event.

 To aid in the interpretation of these light curves we have also
 assembled some images of this event, which are shown in
 Figure~\ref{fig:4}. In addition to the XRT Al\_poly/Open and
 Al\_poly/Al\_thick images we show the STEREO B EUVI Fe\,\textsc{ix/x}
 171\,\AA\ images from this time \citep{howard2002}. There are no
 TRACE images during the event because the spacecraft was experiencing
 an orbital eclipse. These images indicate how complex even a small
 transient event can be. The XRT images show that the emission at high
 temperatures does not come from a single loop but several loops. The
 171\,\AA\ images show an intense footpoint brightening, suggesting
 that the spike in the Si\,\textsc{vii} emission originates in one of
 the loop footpoints.

 The appearance of the loop in the XRT Al\_thick filter suggests that
 the emission reaches very high temperatures. We have found signatures
 of this high temperature emission in the EIS spectra. The
 Fe\,\textsc{xxiv} 255.1 and 192.0\,\AA\ lines appear very briefly in
 these data, peaking at about 15:45, coincident with the peak in the
 GOES 1--8\,\AA\ flux. At this time Fe\,\textsc{xxiii} 263.76\,\AA\ also
 appears. During this event emission at Fe\,\textsc{xvii} 254.87\,\AA\
 and Ca\,\textsc{xvii} 192.87\,\AA\ are also observed. For this
 relatively weak event, the Ca\,\textsc{xvii} emission is difficult to
 disentangle from the Fe\,\textsc{xi} and O\,\textsc{v} blends.

\begin{figure}[t!]
 \centerline{\FigureFile(3in,2.5in){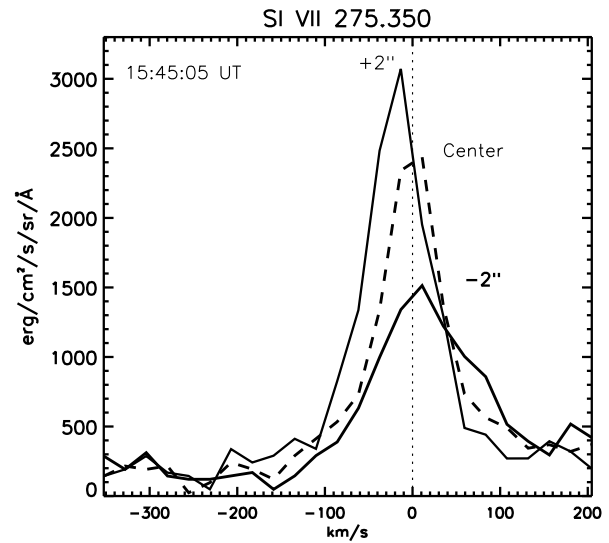}}
 \caption{Line profiles from Si\,\textsc{vii} 275.35\,\AA\ in one of the
 loop footpoints. The line profiles away from the center of the
 brightening show relative Doppler shifts suggesting to formation of a
 reconnection jet.}
 \label{fig:5}
 \end{figure}

 The footpoint emission observed at the lower temperatures (such as
 Si\,\textsc{vii}) have an interesting Doppler signature. Profiles near
 to the top of the brightening show a significant shift relative to
 profiles at the bottom of the brightening. This phenomenon in the
 Si\,\textsc{vii} profile is illustrated in Figure~\ref{fig:5}. Here the
 profile $+2\arcsec$ from the center of the footpoint is blueshifted
 by about 30\,km~s$^{-1}$, but is approximately Gaussian. The profile
 $-2\arcsec$ from the center shows only a small shift in the peak, but
 the profile is broadened out beyond 100\,km~$^{-1}$. This behavior
 suggestive of the bi-direction jets observed in transition region
 emission lines with the SUMER spectrometer on SOHO. These
 bi-directional flows have been interpreted as evidence for magnetic
 reconnection \citep{innes1997}.

\begin{figure}[t!]
 \centerline{\FigureFile(2.4in,3.5in){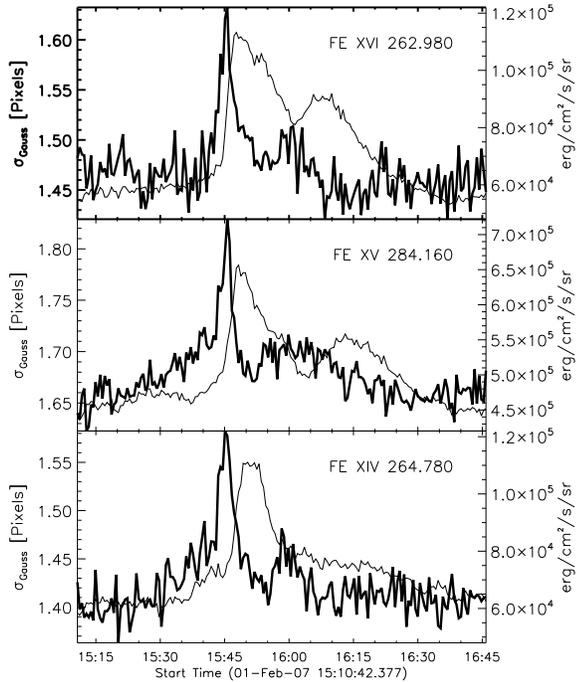}}
 \caption{The evolution of the line width $\sigma_{\rm Gauss}$ (thick
 line) and intensity (thin line) for Fe\,\textsc{xvi} 262.98\,\AA,
 Fe\,\textsc{xv} 285.15\,\AA, and Fe\,\textsc{xiv} 264.78\,\AA. The
 line widths peak before the peak in the line intensity.}
 \label{fig:6}
 \end{figure}

 The emission observed at the higher temperatures with EIS does not
 appear to show significant Doppler signatures. The line centroids are
 generally within a few km~s$^{-1}$ of their pre-flare positions. The
 line widths, in contrast do show some interesting behavior during the
 heating event. As illustrated in Figure~\ref{fig:6}, the line widths
 for Fe\,\textsc{xvi}, Fe\,\textsc{xv}, and Fe\,\textsc{xiv} peak
 during the rise phase in the event. In all of the lines, the peak in
 the line width is achieved before the peak in the line intensity. A
 similar behavior is often observed in flares, where the peak
 non-thermal velocity typically occurs at or before the peak in
 intensity \citep{alexander1998,mariska1999}. In the flare case it has
 been conjectured that this is related to turbulent motions during the
 heating. Since the plasma observed in these emission lines is
 presumably cooling, the origin of the excess width early in the event
 is difficult to interpret.

 \section{Discussion}

 We have presented an initial look at active region transient
 brightenings observed with the EIS and XRT instruments on Hinode. The
 high spatial and temporal resolution of these instruments indicate
 that high temperature active region emission can be highly dynamic,
 suggesting that impulsive heating is more important than previously
 thought. The fine temperature resolution of EIS allows us to follow
 the evolution of active region plasma in great detail. We see
 numerous examples of plasma cooling from high temperatures to lower
 temperatures and have presented one example in some detail. To our
 knowledge this is the most complete observation of an active region
 transient to date. The very broad temperature coverage of EIS allows
 us to follow the temperature evolution from above 10\,MK to below
 0.1\,MK. The high spectral resolution of EIS also allows us to
 identify possible velocity signatures of active region heating in one
 event. We find what appear to be bi-directional flows at the loop
 footpoints and excess line widths during the rise phase of the event.

 These results, however, are preliminary. They generally come from the
 analysis of a single event from a single active region. Our
 impression is that this event is qualitatively similar to the other,
 smaller events that are observed in this region. Many more systematic
 studies of active region transient activity with Hinode are clearly
 required.

 While the analysis of transient events observed with XRT and EIS has
 the potential to yield insights into the coronal heating mechanism,
 it is also possible that there is a strong steady heating level that
 contributes to the active region heating. The light curves presented
 in Figure~\ref{fig:2} clearly show a high basal level of several
 thousand DN~s$^{-1}$. Differentiating between steady heating with
 transient events superimposed and purely impulsive heating will
 require detailed statistical analysis and modeling of the observed
 light curves (e.g., \cite{shimizu1992}). 
 
 The heating of the solar corona is undoubtedly related to the
 evolution of surface magnetic fields and incorporating the analysis
 of magnetic field data and chromospheric emission from the SOT on
 Hinode is also very important. We have surveyed the SOT data from
 this period and we see a wealth of activity at these heights in the
 solar atmosphere. Much of the photospheric flux, however, closes
 before it reaches the corona \citep{close2003} so the challenge is to
 determine what activity is relevant to the coronal dynamics we
 observe.

 \vspace{0.1in} 

Hinode is a Japanese mission developed and launched by ISAS/JAXA, with
NAOJ as domestic partner and NASA and STFC (UK) as international
partners. It is operated by these agencies in co-operation with ESA
and NSC (Norway). We thank Spiros Patsourakos for preparing the STEREO
EUVI data for us. We also thank the referee for a thorough reading of
the manuscript and many insightful comments.


\end{document}